\documentstyle[12pt]{article}
\newcommand{\be}{\begin{equation}}
\newcommand{\ee}{\end{equation}}
\newcommand{\bea}{\begin{eqnarray}}
\newcommand{\eea}{\end{eqnarray}}
\newcommand{\ba}{\begin{array}}
\newcommand{\ea}{\end{array}}

\def\bbox{{\,\lower0.9pt\vbox{\hrule \hbox{\vrule height 0.2 cm
\hskip 0.2 cm \vrule height 0.2 cm}\hrule}\,}}
\newcommand{\dsl}{\pa \kern-0.5em /}
\newcommand{\lb}[1]{\label{#1}}
\newcommand{\p}[1]{(\ref{#1})}

\newcommand{\nn}{\nonumber \\}

\font\mybb=msbm10 at 10pt
\def\bb#1{\hbox{\mybb#1}}

\def\bE {\bb{E}}

\def\bC {\bb{C}}

\def\appendix#1{
  \addtocounter{section}{1}
  \setcounter{equation}{0}
  \renewcommand{\thesection}{\Alph{section}}
  \section*{Appendix \thesection\protect\indent \parbox[t]{11.15cm}
  {#1} }
  \addcontentsline{toc}{section}{Appendix \thesection\ \ \ #1}
  }

\begin{document}



\begin{titlepage}
\rightline{DAMTP-05-88}
\rightline{\tt{hep-th/0510019}}

\vfill

\begin{center}
\baselineskip=16pt {\Large\bf  Planar Super-Landau Models}
\vskip 0.3cm
{\large {\sl }}
\vskip 10.mm
{\bf ~Evgeny Ivanov$^{*,1}$, Luca Mezincescu$^{\dagger,2}$ and  Paul K.
Townsend$^{+,3}$}
\vskip 1cm
{\small
$^*$
Bogoliubov Laboratory of Theoretical Physics \\
JINR, 141980 Dubna, Moscow Region, Russia\\
}
\vspace{6pt}
{\small
$^\dagger$
Department of Physics,\\
University of Miami,\\
Coral Gables, FL 33124, USA\\
}
\vspace{6pt}
{\small
  $^+$
Department of Applied Mathematics and Theoretical Physics\\
Centre for Mathematical Sciences, University of Cambridge\\
Wilberforce Road, Cambridge, CB3 0WA, UK\\
}
\end{center}
\vfill

\par
\begin{center}
{\bf ABSTRACT}
\end{center}
\begin{quote}

In previous papers we solved the Landau problems, indexed by $2M$,
for a particle on the ``superflag''  $SU(2|1)/[U(1)\times U(1)]$,
the $M=0$ case being equivalent to the Landau problem for
a particle on the ``supersphere'' $SU(2|1)/[U(1|1)]$.
Here we solve these models in the planar limit.
For $M=0$ we have a particle on the complex
superplane $\bC^{(1|1)}$; its Hilbert space is the
tensor product of that of the Landau model with
the 4-state space of a ``fermionic'' Landau  model.
Only the lowest level is ghost-free, but for  $M>0$ there are
no ghosts in the first $[2M]+1$ levels. When $2M$ is an integer, the
$(2M+1)$th level states form short supermultiplets as a consequence
of a fermionic gauge invariance analogous to the ``kappa-symmetry''
of the superparticle.

\vfill
\vfill
\vfill
\vfill
  \hrule width 5.cm
\vskip 2.mm
{\small
\noindent $^1$ eivanov@theor.jinr.ru\\
\noindent $^2$ mezincescu@server.physics.miami.edu\\
\noindent $^3$ p.k.townsend@damtp.cam.ac.uk
\\ }
\end{quote}
\end{titlepage}

\section{Introduction}
\setcounter{equation}{0}

In 1930 Landau posed and solved the quantum mechanical problem of a charged
particle in a plane orthogonal to a uniform magnetic field, showing in particular that the
particle's energy is restricted to a series of ``Landau levels'' \cite{landau}.  
In the low-energy limit
only the lowest level is relevant, and the low-energy physics is described by a first-order
``Lowest-Landau-Level'' (LLL) model with a phase space that is a non-commutative version
of the original configuration space. In more recent times, this connection with non-commutative
geometry has led to a revival of interest in Landau-type models.

In 1983 Haldane generalized the Landau model to a particle on a sphere in $\bE^3$
of radius $R$,  in the uniform magnetic field  $B$ generated by a magnetic monopole at the
centre of the sphere \cite{Haldane:1983xm}. If the monopole strength is $n$
times the minimal value allowed by the Dirac quantization condition
then $B\propto n/R^2$ and the planar Landau model is recovered
in the limit that $n\to\infty$ and $R\to\infty$ with $B$ kept fixed.
If instead one takes the limit as $R\to 0$ with $n$ fixed then one finds
a LLL model with an action that is $n$ times the minimal $U(1)$
Wess-Zumino (WZ) term associated with the description of the
sphere as $SU(2)/U(1)\cong CP^1$.
The phase space of this LLL model is a fuzzy sphere  \cite{Madore:1991bw}.

In two previous papers \cite{Ivanov:2003qq,Ivanov:2004yw}
we have considered Landau models for a particle on a {\it superspace} with
$CP^1$ body. The minimal dimension symmetric superspace with this property  is
$CP^{(1|1)} \cong SU(2|1)/U(1|1)$, which we called the
{\it supersphere}\footnote{Other  definitions of  ``supersphere''  occur in the literature
 (references can be found in our previous papers)  but
 none is obviously equivalent to our definition.}. The LLL model
 for a particle on the supersphere yields a physical realization
 of the fuzzy supersphere  \cite{Ivanov:2003qq}.  Although this model is perfectly physical,
 the full Landau model for a particle on the supersphere is unphysical because
 the higher Landau levels all contain ghosts; i.e., states of negative norm.  This feature is
 directly related to the presence of a non-canonical fermionic kinetic term
 with two time derivatives.

In an attempt to circumvent this problem, we considered in  \cite{Ivanov:2004yw}
the Landau model for a particle on  the coset superspace $SU(2|1)/[U(1)\times U(1)]$.
This supermanifold again has $CP^1$ body but it is not a symmetric superspace;
it is an analog of the flag manifold $SU(3)/[U(1)\times U(1)]$, and for this reason we
called it the  ``superflag''.  For given magnetic field strength there is a one-parameter family
of superflag Landau  models  parametrized, in the notation of \cite{Ivanov:2004yw},
by the coefficient $2M$ of an  additional, purely ``fermionic'', WZ term.
Although the relationship between the superflag and supersphere Landau models
was not spelled out in our earlier work, one can show that supersphere model is equivalent
to the $M=0$ superflag model. The parameter $M$ has no effect
on the energy levels, which are therefore the same as those of the supersphere
Landau model, but  one now finds that states in the first $[2M]+1$ levels have positive norm,
although all higher  levels still contain states of negative norm\footnote{
$[2M]$ is the integer part of $2M$.}.  When $2M$ is an integer, the $(2M+1)$th level states
form a short representation of $SU(2|1)$ as a consequence of the presence
of zero-norm states.

One aim of this paper is to elucidate these features of spherical super-Landau models
by an analysis of the much simpler models obtained in the planar limit.
The planar limit of the supersphere is the complex superplane $\bC^{(1|1)}$. This
 can be viewed as the coset superspace
\be
IU(1|1)/[U(1|1) \times {\cal Z}]\,,
\ee
where $IU(1,1)$ is a central extension of a contraction of $SU(2|1)$ and
${\cal Z}$ is the abelian group generated by the central charge.
The corresponding ``superplane Landau model''  has a quadratic Lagrangian
and a Hilbert space
that is the  tensor product of the standard Landau model Hilbert space with a 4-state
space of  a ``fermionic Landau model''.  Analysis of this 4-state system shows clearly
how negative norm states arise as a consequence of the two-derivative,
and hence non-canonical,  fermion kinetic term, but also why the LLL is ghost-free.

This analysis of the superplane Landau model suggests a strategy for
removing the negative norm states by modifying the Lagrangian in such a
way as to cancel the two-derivative, or ``second-order'',  fermion kinetic term.
This requires the introduction of interactions with an additional
complex ``Goldstino'' variable and the
introduction of a first-order  kinetic term for it, with coefficient $2M$.
The resulting model, which is
the Landau model for a particle on the coset superspace
\be
IU(1|1)/[U(1)\times U(1) \times {\cal Z]}\, ,
\ee
is precisely the planar limit of the superflag Landau model;
we call it the ``planar superflag Landau model''. The cancellation
of the second-order  fermion term in this  ``planar superflag''  Landau
model is incomplete, however, because it survives in a ``bodyless'' form
with nilpotent Goldstino bilinear coefficient. At the quantum level, this results
not in the elimination of all negative norm states but rather in their
banishment to the higher Landau levels, exactly as found in \cite{Ivanov:2004yw}
for the superflag Landau model. One may then discard these levels to arrive at a 
model with a finite-dimensional Hilbert space that generalizes the LLL model 
obtained by the truncation to the ground state level, exactly as argued in 
\cite{Ivanov:2004yw} for the superflag Landau model.

Thus many of the peculiar properties of
the supersphere and superflag models of  \cite{Ivanov:2003qq,Ivanov:2004yw}
survive the planar limit and are readily understood in this simpler context.
In particular, the structure of the phase-space constraints is simple to analyse
in the planar limit, and it explains why zero norm states appear in the $(2M+1)$th level
when $2M$ is an integer. Recall that the
Hamiltonian formulation of models with canonical fermion kinetic terms requires
fermionic constraints on the phase superspace. No such constraints are needed for
the superplane model as it has non-canonical, second-order, fermion kinetic terms,
but constraints are needed for the ($M>0$) planar superflag model. Usually, these
constraints are either  all ``second class'' (in Dirac's terminology) or (as in many
superparticle models) a definite mixture of first and second class, the first class constraints
indicating the presence of a fermionic gauge invariance. Here we find fermionic constraints
that are second class everywhere {\it except on a particular energy surface}, where they
are of mixed type\footnote{Something similar occurs for higher-dimensional
Chern-Simons theories \cite{Banados:1996yj} but in the context of bosonic constraints.}.
This implies a fermionic gauge invariance analogous to the ``kappa-symmetry''
of the superparticle, but restricted to a subspace of definite energy.
Because of energy  quantization, this has an effect on the quantum theory
only when $2M$ is an integer, and it is responsible for the zero-norm states
in the $(2M+1)$th level.

We shall begin with an analysis of the superplane Landau model. Its quantization is
essentially trivial because the Lagrangian is quadratic, but it provides a useful
starting point, and a simple context in which one can discuss the $IU(1|1)$ symmetries.
We then show how a modification of this Lagrangian to include
interactions with a Goldstino variable yields the planar superflag Landau models,
indexed by the coefficient $2M$ of a fermionic WZ term.
The equivalence with the superplane Landau model
for $M=0$ is then established; this equivalence is not obvious and requires careful
consideration of the Hamiltonian constraint structure of the planar superflag models.
We then use this Hamiltonian analysis to quantize the planar superflag model, using
the methods of our previous papers. Finally, we present a geometrical
formulation of our results.

\section{The Superplane Landau Model}
\setcounter{equation}{0}

We begin with the superplane Landau model. By ``superplane'' we mean the superspace
$\bC^{(1|1)}$ parametrized by complex coordinates $(z,\zeta)$,
where $z$ is a complex number and $\zeta$ a complex  anticommuting variable.
The superplane Lagrangian is
\be
L_0= L_b + L_f\,, \label{L0}
\ee
where
\be
L_b = |\dot z|^2 -i\kappa \left(\dot z \bar z - \dot{\bar z} z\right)\label{Bos}
\ee
is the Lagrangian for the standard planar Landau model with energy spacing $2\kappa$
(which we take to be positive), and
\be
L_f = \dot\zeta\dot{\bar\zeta} -i\kappa \left(\dot\zeta\bar\zeta 
+ \dot{\bar\zeta} \zeta\right) \label{Ferm}
\ee
is the Lagrangian for a fermionic Landau model in terms of an anticommuting complex variable
$\zeta$.  We call the total Lagrangian $L_0$ because it is quadratic;
we will later add interaction terms to get the Lagrangian of the planar superflag Landau model.

The Hilbert space of this model is obviously a tensor product of the Hilbert space
of the Landau model with that of the fermionic Landau model with Lagrangian $L_f$,
so all the new features must arise from the latter model, which we therefore analyse first.
Because $L_f$ contains a ``second-order'' kinetic
term, and second-order is ``higher-order'' for fermions,  we should expect ghosts
(negative norm states). We shall show that this intuition is indeed correct, but
also  that all LLL levels have positive norm. This too is expected since the LLL states are all
that survive in large $\kappa$ limit in which all terms of the second order in time derivative
become irrelevant.

Having analysed the fermionic Landau model, the spectrum of states of the full superplane model,
and their norms, is easily determined. However, the degeneracies in the spectrum are consequences
of symmetries of the full Lagrangian. The relevant symmetry group is the 
supergroup $IU(1|1)$ obtained
by a contraction of the $SU(2|1)$ symmetry of the supersphere. We exhibit these symmetries, and
show precisely how  $IU(1|1)$ is obtained from $SU(2|1)$.

\subsection{Fermionic Landau model}

For the purposes of comparison we first summarize Landau's results for the standard,
``bosonic'' Landau model. The equation of motion has the general solution
\be
z= z_0 + (\dot z_0/\kappa)e^{-i\kappa t} \sin \kappa t\, ,\label{a}
\ee
so the motion is periodic with angular frequency $2\kappa$. To pass
to the quantum theory it is convenient to use the Hamiltonian form of the Lagrangian
\be
L_b = \dot z p + \dot{\bar z}\bar p - H_b\,, \qquad 
H_b= \left|p + i\kappa \bar z\right|^2\, ,
\ee
where $p$ is the complex momentum conjugate to $z$. To obtain the quantum Hamiltonian
$\hat H_b$ we then make the replacements
\be
p \to \hat p = -i\partial_z \,, \qquad \bar p \to \hat {\bar p} 
= -i\partial_{\bar z}\, .
\ee
There is a trivial ordering ambiguity but the natural symmetric ordering yields
\be
\hat H_b = a^\dagger a + \kappa\,, \label{Bhamilt}
\ee
where
\be
a =  i\left(\partial_{\bar z} + \kappa z\right) \, , \qquad
a^\dagger = i\left(\partial_z - \kappa \bar z\right).
\ee
These operators satisfy the creation and annihilation operator commutation relation
\be
[a,a^\dagger] = 2\kappa\,.
\ee
The ground states, which span the LLL, have energy $\kappa$ and are annihilated by $a$.
States in the higher Landau levels are obtained by acting on a LLL state with $a^\dagger$,
so the energy levels are $E= 2\kappa(N+1/2)$ for non-negative integer $N$.

The equation of motion of the fermionic Landau model has the general solution
\be
\zeta = \zeta_0 +  (\dot \zeta_0/\kappa)e^{-i\kappa t} \sin \kappa t\,, \label{b}
\ee
so the motion is again periodic with period $2\kappa$. The Hamiltonian form of the Lagrangian is
\be
L_f = -i \dot \zeta \pi - i\dot{\bar\zeta}\bar\pi  -H_f\, ,\qquad H_f =  
\left(\bar\pi - \kappa\zeta\right)
\left(\pi - \kappa \bar\zeta\right)\, ,
\ee
where $\pi$ is the momentum conjugate to $\zeta$. We use here the Grassmann-odd phase
space conventions of \cite{Ivanov:2003ax}  for which $\bar\pi$ is the complex conjugate of $\pi$.
Note that this Lagrangian is invariant under the rotational and translational
isometries of the complex Grassmann plane (together with a corresponding phase rotation of $\pi$).
To pass to the quantum theory we make the replacements
\be\label{replaceferm}
\pi \to \hat \pi = \partial_\zeta \,, \qquad 
\bar\pi \to \hat{\bar\pi} = \partial_{\bar\zeta}\, ,
\ee
where the Grassmann-odd derivatives should be understood as left-derivatives.
There is a trivial ordering ambiguity in the Hamiltonian, but the natural
antisymmetric ordering yields the quantum Hamiltonian\footnote{By changing the sign of $\kappa$
and interchanging the roles of annihilation and creation operators, this could be brought
to the form $H= \alpha^\dagger \alpha -\kappa$, which is the standard form 
for a fermionic oscillator. However, the formulation given here is 
the one most convenient for the purpose of combining
it with the standard Landau model to get the superplane Landau model that we consider here.}
\be\label{fermham}
\hat H_f =  - \alpha^\dagger \alpha - \kappa\,,
\ee
where
\be
\alpha= \left(\partial_{\bar\zeta} - \kappa \zeta\right)\, , \qquad
\alpha^\dagger = \left(\partial_\zeta - \kappa \bar\zeta\right)\, .
\ee
These operators satisfy the anticommutation relations
\be
\{\alpha, \alpha^\dagger\} = -2\kappa\, .
\ee
The Hamiltonian $\hat H_f$ has four linear independent eigenfunctions $\Psi(\zeta,\bar\zeta)$.
Two, which we denote collectively by $\Psi_-$, have energy $-\kappa$ and the other two,
which we denote collectively by $\Psi_+$, have energy $+\kappa$. From the requirement
that $\Psi_-$ is annihilated by $\alpha$ and $\Psi_+$ is annihilated by $\alpha^\dagger$,
it can be seen that
\bea
\Psi_{-} &=& A_- \left(1 + \kappa \bar\zeta \zeta\right) + B_-\zeta\,, \nonumber\\
\Psi_{+} &=& A_+ \left(1 - \kappa \bar\zeta \zeta\right) + B_+\bar\zeta\, .
\eea
Note that $\Psi_+$ can be viewed as an excited state generated by 
the creation operator $\alpha^\dagger$
from the ``vacuum'' $\Psi_{-}$.

Up to an overall factor, which we may choose at our convenience, the natural inner product
on wavefunctions (invariant under translations and phase rotations of $\zeta$) is
\be
\langle  \Psi_1,\Psi_2\rangle  = \partial_\zeta \partial_{\bar\zeta} 
\left(\Psi_1^* \Psi_2 \right)\, .
\ee
It is straightforward to verify that wavefunctions with different energies are orthogonal with
respect to this inner product, and that
\bea
\langle \Psi_-,\Psi_-\rangle &=&
2\kappa \bar A_- A_- + \bar B_- B_-\, ,  \nonumber\\
\langle \Psi_{+},\Psi_{+}\rangle &=&
-2\kappa \bar A_{+} A_{+} - \bar B_{+} B_{+}\, .
\eea
In arriving at this result we have been careful not to assume any particular Grassmann-parity
for the complex constants $A$ and $B\,$. It would be possible to suppose 
that all are Grassmann even,
in which case it is clear that if the states of $\Psi_-$ have positive norm 
then the states of $\Psi_+$
have negative norm. If instead one assumes that $\Psi_-$ and $\Psi_{+}$ have 
a definite Grassmann parity,
so that either the $A$ or the $B$ coefficient is Grassmann-odd, then it is still 
true that the states
of $\Psi_-$ have non-negative norm (this now being a complex supernumber) 
while those of the higher
level have a non-positive norm, with some state of negative norm, 
and this is true whatever
assumption one makes about the Grassmann parity of  $\Psi_\pm\,$. Thus, 
{\it only $\Psi_-$ has all states of non-negative norm}.

As for the standard Landau model, one can take a limit in which only 
the lowest Landau level survives.
The corresponding LLL Lagrangian is just the fermion WZ term. 
This is the simplest case
of the ``odd-coset'' models studied in \cite{Ivanov:2003ax}, with a Hilbert 
space spanned by the two positive-norm states.

Before moving on, we pause to comment on the limit in which $\kappa=0$. The bosonic Landau
model becomes a model for a particle moving freely on the complex plane. 
In contrast, the fermionic Landau model  is unphysical when $\kappa=0$ 
because the Hamiltonian operator $\hat H_f$ is then nilpotent 
and hence non-diagonalizable. For this reason we henceforth consider only $\kappa\ne0$. 
Although this restriction is unphysical in the Landau model, 
where $\kappa$ is proportional to the magnetic field, it may be physical 
in any context in which the fermionic Landau model plays 
a role since the parameter $\kappa$ may then have some other interpretation.

\subsection{The superplane model and its symmetries}

We now return to the Landau model for a particle on the superplane. 
The Hamiltonian form
of the Lagrangian is
\be
L_0= \left(\dot z p -i \dot \zeta \pi\right) + c.c. - (H_b + H_f)\,. \label{L0ham}
\ee
The quantum Hamiltonian has energy levels $2\kappa N$ for non-negative integer $N$. 
In particular
the states $|LLL\rangle$ of the LLL have zero energy and satisfy
\be
a|LLL\rangle = 0 \, ,\qquad \alpha |LLL\rangle =0\, .
\ee
All these states have positive norm. The first exited states (with energy $2\kappa$)
are linear combinations of states of the form $a^\dagger |LLL\rangle$, which
all have positive norm, and
states of the form $\alpha^\dagger|LLL\rangle$, some of which have negative norm.
Thus, only the LLL has all states of positive norm.

Note that the zero point energy cancels between the bosonic and fermionic sectors,
as happens in supersymmetric quantum mechanics. However, the  ``supersymmetry''  of the
superplane Landau model is rather different from that of supersymmetric quantum
mechanics. As for any quadratic Lagrangian (except those with {\it only} Grassmann-odd
variables \cite{Ivanov:2003ax}), the full symmetry
group is infinite-dimensional. However, the symmetries of relevance here are those
inherited from the supersphere. These are the super-translations of the superplane,
the $SU(1|1)$ super-rotations,
and an independent $U(1)$ phase rotation.

The super-translation transformations are
\be
\pmatrix{ \delta z \cr \delta \zeta} = \pmatrix{c\cr\gamma} \, , \qquad
\pmatrix{\delta p \cr \delta \pi} = \kappa\pmatrix{-i\bar c\cr \bar\gamma}\, ,
\ee
for complex constant $c$ and complex Grassmann-odd constant  $\gamma$. This symmetry
is generated by the operators
\bea
P &=&  -i\left(\partial_z + \kappa \bar z\right)\, ,\qquad
P^\dagger = -i\left(\partial_{\bar z} -\kappa z\right)\nonumber \\
\Pi &=& \partial_\zeta + \kappa \bar\zeta\, , \qquad
\Pi^\dagger = \partial_{\bar\zeta} + \kappa \zeta\, . \label{Pnoncomm}
\eea
Their non-zero (anti)commutation relations are
\be
[P, P^\dagger] = 2\kappa \, ,\qquad \{\Pi^\dagger,\Pi\} = 2\kappa\, .\label{100}
\ee
Thus, $\kappa$ is a central charge. We will call the superalgebra 
defined by these relations
the ``magnetic translation superalgebra''.

The  $SU(1|1)$ super-rotation transformations are
\be
\pmatrix{ \delta z \cr \delta \zeta} =
 \pmatrix{i\theta & -\bar\epsilon \cr -\epsilon & i\theta}
\pmatrix{z \cr \zeta}\, , \qquad
\pmatrix{\delta p\cr \delta \pi} = \pmatrix{-i\theta & -i\epsilon \cr
-i\bar\epsilon & -i\theta} \pmatrix{p \cr  \pi}
\ee
for constant angle $\theta$ and complex Grassmann-odd  parameter $\epsilon$.
The odd transformations are generated by the operators
\be
Q =  z \partial_\zeta - \bar\zeta\partial_{\bar z}\, , \qquad
Q^\dagger = \bar z \partial_{\bar\zeta} + \zeta\partial_z \label{QQbar}
\ee
and the even transformation is generated by the Hermitian operator
\be
C= z\partial_z + \zeta\partial_\zeta - \bar z\partial_{\bar z}
- \bar\zeta\partial_{\bar\zeta}\, . \label{Ccharge}
\ee
The only non-zero  (anti)commutation relations of these generators is
\be
\{Q,Q^\dagger\} = C\, .\label{101}
\ee
This is analogous to a standard supersymmetry algebra but with $C$ as the Hamiltonian.
It should be noted, however, that many of the usual consequences of supersymmetry
would not apply anyway because of the negative-norm states.

The $SU(1|1)$ charges, together with the supertranslation charges,
span a semi-direct product superalgebra which we will call $ISU(1|1)$. In particular,
\be
[Q,P] =i \Pi\, , \qquad \{Q^\dagger,\Pi \} = iP\,, \qquad [C,P]=-P\, ,
\qquad [C,\Pi]=-\Pi\, .\label{102}
\ee
However, as shown by (\ref{100}), we must include a central charge $Z=\kappa$;
this generates an abelian group, which we call ${\cal Z}$ and
include as part of the definition of $ISU(1|1)$. The superplane can now be viewed
as the coset  superspace $ISU(1|1)/ [SU(1|1)\times {\cal Z}]\,$.

Finally we have an independent $U(1)$ phase rotation with infinitesimal transformations
\bea
\delta z &=& i\varphi z\, ,\qquad \delta p = -i\varphi p\, , \nonumber\\
\delta \zeta &=& -i\varphi \zeta \, ,\qquad \delta \pi = i\varphi \pi\, .
\eea
This is generated by the Hermitian operator
\be
J=  {1\over2}\left[z\partial_z - \zeta\partial_\zeta - \bar z\partial_{\bar z}
+ \bar\zeta\partial_{\bar\zeta}\right] \label{Jdef}
\ee
which has the following non-zero commutation relations with the generators 
of $ISU(1|1)$
\be
[J,Q] = Q\, ,\qquad [J,Q^\dagger] = -Q^\dagger\, , \qquad [J,P]=-P\, ,
\qquad [J,\Pi] = \Pi\, . \label{Jcomm}
\ee
The supergroup generated by the five even charges $(P,P^\dagger, C,J, Z)$
and the four odd charges $(\Pi, \Pi^\dagger, Q, Q^\dagger)$ will
be called $IU(1|1)$, and the superplane can be viewed as the coset superspace
$IU(1|1)/[U(1|1)\times {\cal Z}]$, as mentioned in the introduction.
This has the advantage that $IU(1|1)$ is a contraction of $SU(2|1)$, as we now show.

\subsection{IU(1$|$1) as contraction of SU(2$|$1)}

We now sketch how the algebra of the supergroup $IU(1|1)$ defined by the
relations \p{100}, \p{101}, \p{102} and \p{Jcomm} can be reproduced as a contraction
of the superalgebra $su(2|1)$. The contraction procedure is similar to the one relating
$su(2)$ to the algebra of magnetic translations \cite{Hatsuda:2003ry}.

The bosonic body of the superalgebra $su(2|1)$ is $su(2)\oplus u(1)$
with the generators $J_{\pm}, J_3$ and $B$ \cite{Ivanov:2004yw}
\bea && [J_+,
J_-] = -J_3\,, \; [J_3, J_\pm ] = \pm 2J_\pm\,, \quad [B, J_3] =
0\,, \; [B, J_\pm] = \mp J_\pm\,, \nn && J_3^\dagger = J_3\,,
\;B^\dagger = B\,, \; J_+^\dagger = -J_-\,.
\eea
The odd sector is spanned by an $SU(2)$ doublet generators $S_1, S_2, \bar S^1, \bar
S^2$ with the following non-vanishing (anti)commutation relations
(and their conjugates):
\bea
&& \{S_1, \bar S^1\} = J_3 + B\,, \;
\{S_2, \bar S^2\} = B\,, \; \{S_1, \bar S^2\} = -J_+\,, \; \{S_2,
\bar S^1\} = J_-\,, \nn
&& [J_3, S_1] = S_1\,, \; [J_3, S_2] = -
S_2\,, \; [B, S_1] = -Q_1\,, \; [B, S_2] = 0\,, \nn && [J_+, S_1] =
0\,, \; [J_+, S_2] = -S_1\,, \;[J_-, S_1] = S_2\,, \; [J_-, S_2] =
0\,. \label{sl21}
\eea
Note that the second $U(1)$ generator $B$ basically has
the same commutation relations with $J_\pm$ as $J_3$,\footnote{This set of generators can
be split into the mutually commuting $u(1)$ and $su(2)$ sets by passing to the
appropriate linear combination of $B$ and $J_3$, but we prefer to use this basis in order to
have a correspondence with the notation of ref. \cite{Ivanov:2004yw}.}
but both these generators ($B$ and $J_3$) have different
action on the spinors.

The contraction leading to the magnetic translation superalgebra introduced in the
previous subsection  goes as follows. Firstly one redefines
(and/or rename) the generators as
\be
J_3 = 2n - 2J\,,\; J_+ =
iR\,P\,, \; J_- = iR\,P^\dagger\,, \; S_1 = R\,\Pi\,, \; S_2 =
-Q\,, \; B =C\,,
\ee
where $n$ and $R$ are two real parameters ($R$ is a
radius of the sphere $S^2 \sim SU(2)/U(1)$ while $n$, in the
dynamical framework of a particle moving on the superflag manifold
$SU(2|1)/[U(1)\times U(1)]$ \cite{Ivanov:2004yw}, acquires a nice meaning of
the strength of the $SU(2)/U(1)$ WZW term). Then one substitutes
this into \p{sl21} and let $R \rightarrow \infty\,$, assuming that
\be
\frac{n}{R^2} \equiv \kappa < \infty\,.
\ee
As the result of this contraction procedure, the algebra of the $su(2)$ generators
$J_\pm, J_3$ in \p{sl21} goes over into the magnetic translation
algebra (given by the first relation in \p{100}) and the relations \p{sl21}
become just \p{101}, \p{102} and the second relation in \p{100} (plus the
evident additional commutation relations with the generator $J\,$, eq. \p{Jcomm}).
It is worth noting that, in the contraction limit, one of the $U(1)$
charges, $J$, fully decouples and generates an outer $U(1)$
automorphism, while $B \equiv C$ still remains in the r.h.s. of
$\{S_2, \bar S^2 \}\,$. Another notable feature is the appearance
of the constant central charge $\kappa$ which thus formally extends the full
number of bosonic generators to five as compared with four such generators
in $SU(2|1)$; this also happens in the purely bosonic $su(2)$ or $sl(2,R)$
cases \cite{Hatsuda:2003ry}.

\section{The Planar Superflag Landau Model}
\setcounter{equation}{0}

The problem with the Landau model on the superplane is that the second-order Lagrangian
for the Grassmann-odd variable implies the presence of ghosts (negative norm states)
in the quantum theory. This is forced by the $Q$-supersymmetry of $SU(1|1)$ that
relates bosons to fermions, so any solution to this problem would appear
to require a breaking of this symmetry, but we would need the breaking
to be spontaneous in order to maintain the $IU(1|1)$ symmetry of the Lagrangian.
This suggests that we aim for a non-linear realization of the $Q$-supersymmetry
by introducing a Goldstino variable $\xi$ with
the $Q$-transformation
\be
\delta \xi = \epsilon\, .
\ee
We now observe that the new Lagrangian
\be\label{altlag}
\tilde{L} = L_0  -   \left(|\dot z|^2 + \dot\zeta \dot{\bar\zeta}\right)
\left(\xi + \dot\zeta/\dot z\right)
\left(\bar\xi + \dot{\bar\zeta}/\dot {\bar z}\right)
\ee
is invariant under all the symmetries previously established for $L_0$.
Collecting terms, we have
\be
\tilde{L} = \left(1+\bar\xi\xi\right)|\dot z|^2 + \left(\bar\xi \dot{\bar z} \dot \zeta
- \xi \dot z \dot{\bar\zeta}\right) + \bar\xi\xi \dot\zeta \dot{\bar\zeta}
- i\kappa\left(\dot z \bar z - \dot{\bar z} z+ \dot\zeta\bar\zeta + \dot{\bar\zeta} \zeta\right)\,,
\label{Mod}
\ee
which shows both that the new Lagrangian is well-defined at $\dot z=0$, despite initial
appearances, {\it and} that the second-order kinetic term $\dot\zeta \dot{\bar\zeta}$ term
now has a nilpotent coefficient. The implications of this are not immediately apparent
but will become clear in due course.

Although it might appear that we have now solved, or at least ameliorated, the problem of ghosts,
we have actually just  hidden it; the $\xi$ equation of motion is
\be\label{EqKsi}
\left(\dot{z}\dot{\bar z} + \dot\zeta\dot{\bar\zeta}\right)\xi + \dot{\bar z}\dot\zeta = 0
\ee
and if $\dot z\ne0$ this implies
\be\label{Ksi}
\xi = -\frac{\dot\zeta}{\dot{z}}\, .
\ee
Back substitution into $\tilde L$ yields the quadratic Lagrangian $L_0$ with which we started, so
$\tilde L$ is classically equivalent to $L_0$, except possibly when $\dot z=0$,
which implies zero classical energy. Thus, apart from this subtlety,
to which we return later, nothing has yet been accomplished. However,
there is now an additional WZ term that we can add to the Lagrangian arising
from the closed invariant 2-form $d\bar\xi \wedge d\xi$. This leads us to the Lagrangian
\bea\label{planarSF}
L &=& \left(1+\bar\xi\xi\right)|\dot z|^2 + \left(\bar\xi \dot{\bar z} \dot \zeta
- \xi \dot z \dot{\bar\zeta}\right) + \bar\xi\xi \dot\zeta \dot{\bar\zeta}  \nonumber\\
&&- \ i\kappa\left(\dot z \bar z - \dot{\bar z} z+ \dot\zeta\bar\zeta 
+ \dot{\bar\zeta} \zeta\right)
+iM\left(\bar\xi\dot\xi + \xi\dot{\bar\xi}\right)
\eea
for some constant $M\,$. This model is actually the planar limit of the superflag
Landau model of \cite{Ivanov:2004yw}.

We now proceed to a detailed analysis of this model, in its Hamiltonian formulation,
first classically and then quantum-mechanically. We then provide
a more geometrical derivation of our results based on
the theory of  non-linear realizations.

\subsection{Hamiltonian analysis}

Introducing the complex Grassmann-odd momentum $\chi$ conjugate to $\xi$, the Hamiltonian
form of the Lagrangian (\ref{planarSF}) is\footnote{We here denote by $\tilde p$
the momentum conjugate to $z$ to distinguish it from the momentum conjugate to $z$
in a different set of variables  that we will later use to quantize the model.}
\be\label{lagsf}
L= \left[\dot z\tilde p  -i \dot \zeta \pi  -i \dot\xi \chi  + \lambda_\zeta\varphi_\zeta +
\lambda_\xi \varphi_\xi\right] + c.c. -H\, ,
\ee
where the Hamiltonian is
\be\label{ham1}
H = \left(1- \bar\xi\xi\right) \left| \tilde p +i\kappa \bar z\right|^2
\ee
and the complex Grassmann-odd variables $\lambda_\zeta$ and $\lambda_\xi$
are Lagrange multipliers for the ``fermionic'' constraints $\varphi_\zeta\approx 0$
and $\varphi_\xi \approx 0$ (in Dirac's  ``weak equality'' notation).
The constraint functions are
\be
\varphi_\zeta = \pi -\kappa\bar\zeta + i\bar\xi \left(\tilde p+i\kappa \bar z\right)\, ,\qquad
\varphi_\xi = \chi -M\bar\xi\,.
\ee
To establish the equivalence of (\ref{lagsf}) to (\ref{planarSF}) we solve the constraints
to reduce (\ref{lagsf}) to
\bea\label{lagsf2}
L&=& \left\{\left[\dot z \tilde{p} -i \kappa\, \dot\zeta \bar\zeta -iM\dot\xi \bar\xi\right] 
+ c.c\right\}
-\,  \left| \tilde{p}+i\kappa \bar z\right|^2 -  \dot{\bar\zeta}\dot\zeta   \nonumber\\
&& + \left[ \left(\tilde{p}+i\kappa\bar z\right)\bar\xi + \dot{\bar\zeta}\right]
\left[ \left(\bar{\tilde{p}}-i\kappa  z\right)\xi + \dot\zeta \right].
\eea
Elimination of $\tilde p$ now yields (\ref{planarSF}).

The occurrence of fermion constraints is to be expected in any model with canonical, 
first-order,
fermion kinetic terms, and these constraints are normally second class, 
in Dirac's terminology.
Here, however, we have an additional  ``bodyless''  second-order fermion kinetic term,
and this has a curious consequence. A computation shows that although the Poisson bracket
of the analytic constraint functions $(\varphi_\zeta,\varphi_\xi )$ is zero,
the matrix of Poisson brackets of these functions
with their complex conjugates is non-zero. In fact,
\be
\det \pmatrix{ \{\varphi_\zeta,\bar\varphi_\zeta\}_{PB} 
& \{\varphi_\zeta,\bar\varphi_\xi\}_{PB} \cr
\{\varphi_\xi,\bar\varphi_\zeta\}_{PB}  & \{\varphi_\xi,\bar\varphi_\xi\}_{PB} }
= \left(1+\bar\xi\xi\right) \left[ H-4\kappa M\right].
\ee
It follows that the constraints {\it considered together with their complex conjugates}
are second class everywhere {\it except on the surface $H=4\kappa M$};
on this surface there are first  class constraints.

This unusual state of affairs merits a more detailed analysis. We begin
with the  $M=0$ case, for which the energy surface $H=4\kappa M$ reduces to
the point $H=0$.  As long as the classical energy $(1-\bar\xi\xi)\left| 
\tilde{p}+i\kappa \bar z\right|^2$
(and hence $\left| \tilde{p}+i\kappa \bar z\right|^2$) 
is non-zero we may treat $\xi$ in (\ref{lagsf2})
as an auxiliary variable that can be eliminated by its equation of motion
\be
(\tilde{p} + i\kappa \bar z)\left[(\bar{\tilde{p}} 
- i\kappa z)\xi +\dot\zeta\right] = 0\,.
\ee
This is equivalent to
\be
\xi = -\dot\zeta/\left(\bar{\tilde{p}} - i\kappa z\right)
\ee
provided that $\left|\tilde{p} + i\kappa \bar z\right|^2\neq 0$. 
After substitution for $\xi$
in (\ref{lagsf2}), and subsequent elimination of the momentum variable $\tilde{p}$,
we recover the Lagrangian of the
superplane Landau model. This confirms our analysis of the previous subsection,
but now it is clear how to proceed when the classical energy vanishes;
in this case $\tilde{p} = -i\kappa\bar z$ and
the Lagrangian (\ref{lagsf2}) becomes\footnote{Note that the variables $(z,\zeta)$
are still independent and off-shell.}
\be
L_0= -i\kappa\left\{ \dot z \bar z - z\dot{\bar z}
+ \dot\zeta \bar\zeta + \dot{\bar\zeta}\zeta \right\}.
\label{LLLred}
\ee
This  is the LLL Lagrangian for a particle on the superplane;
the proof of the equivalence of the
superplane model to the $M=0$ planar superflag model is thus completed.

Let us now consider the case of arbitrary $M$. The properties of our model 
on the exceptional
energy surface $H=4\kappa M$ can be studied via a new Lagrangian obtained
by imposing $H=4\kappa M$ as a new, bosonic, constraint 
via a new Lagrange multiplier variable $e(t)$.
The resulting Lagrangian is equivalent to
\bea\label{lagsf3}
L &=& \left[\dot z\tilde p  -i \dot \zeta \pi  -i \dot\xi \chi  
+ \lambda_\zeta\varphi_\zeta +
\lambda_\xi \varphi_\xi\right] + c.c. -  4\kappa M \nonumber\\
&& - \ e\left[ \left|\tilde{p}+i\kappa \bar z\right|^2 
- 4\left(1 +\bar\xi\xi\right)\,\kappa M\right].
\eea
This action is time-reparametrization invariant, with $e$ as the einbein. 
Moreover,
as should be clear from its construction, this action also 
has a hidden fermionic gauge
invariance. In this respect, it is analogous to the superparticle 
action with its hidden
``kappa-symmetry'', the constraint $H=4\kappa M$ being analogous to the standard
mass-shell  superparticle condition with $2\sqrt{\kappa M}$ as a ``mass''.
Many methods have been developed to deal with the mixed first
and second class fermionic constraints of the superparticle, 
and these could be applied here.
Perhaps the simplest is just to solve all the constraints 
to obtain a physical phase-space Lagrangian,
and that is what we will do here.

The fermionic constraints are trivially solved for 
the fermionic momenta $(\pi,\chi)$.
The new bosonic constraint $H= 4\kappa M$ has the general solution
\be
\tilde{p} +i\kappa \bar z =2 e^{i\phi} 
\left(1 +\frac{1}{2}\bar\xi\xi\right)\,\sqrt{\kappa M}\,,
\label{GenSol}
\ee
for some arbitrary phase $\phi(t)$. Using this to eliminate $\tilde p$ 
in favour of $\phi$, we arrive at the
Lagrangian
\bea\label{ConstrLM}
L_{4\kappa M} &=& -i\kappa\left(\dot z \bar z - z\dot{\bar z} +
\dot\zeta\bar\zeta - \zeta\dot{\bar\zeta}\right)
+2\,\left(1 +\frac{1}{2}\bar\xi\xi\right)\,\sqrt{\kappa M}\left[
e^{i\phi} \left(\dot{z} + \bar\xi\dot{\zeta}\right) + c.c.\right]\nonumber \\
&& +\, iM \left(\bar\xi\dot\xi + \xi\dot{\bar\xi} \right) -4\kappa M.
\eea
The new phase variable $\phi$ is actually a gauge variable for 
the $U(1)$ gauge invariance
with infinitesimal gauge transformations
\be\label{AddGauge}
\delta \phi = a(t)\,, \qquad
\delta z = \sqrt{\frac{M}{\kappa}}\left(1+ {1\over2}\bar\xi\xi\right) e^{-i\phi}\, a(t)\,, 
\qquad
\delta \zeta = -\sqrt{\frac{M}{\kappa}} \, \xi e^{-i\phi}\, a(t)\, ,
\ee
where $a(t)$ is the $U(1)$ gauge parameter. This gauge invariance 
allows us to set $\phi(t)=0\,$.
Much more remarkable is the  {\it fermionic} gauge 
invariance with  infinitesimal gauge transformations
\be\label{Kappa}
\delta \xi = \omega\,, \quad \delta \zeta = -i\sqrt{\frac{M}{\kappa}} e^{-i\phi}\omega\,,
\qquad
 \delta z = \frac{i}{2}\sqrt{\frac{M}{\kappa}}
 e^{-i\phi}\left(\bar\omega\xi + \bar\xi\omega\right)\, ,
\ee
where $\omega(t)$ is the complex anticommuting gauge parameter. This gauge invariance
allows us to set $\xi(t)=0\,$.

For the gauge choices $\phi=0$ and $\xi=0\,$, the Lagrangian
\p{ConstrLM} reduces to
\be
L_{4\kappa M} =-i\kappa\left(\dot y \bar y - y\dot{\bar y} +
\dot\zeta\bar\zeta - \zeta\dot{\bar\zeta}\right) - 4\kappa M\,,\label{LLLred2}
\ee
where
\be
y = z -i \sqrt{M/\kappa}\, .
\ee
This is again the LLL Lagrangian for the superplane model, as in  \p{LLLred},
but with the vacuum energy shifted by  $4\kappa M$. We shall see later
that this result has interesting consequences for the quantum theory when $M$
is an integer.

Before turning to the quantum theory we must address a further technical problem;
the Poisson bracket of the Hamiltonian (\ref{ham1}) with the constraint 
function $\varphi_\zeta$
is not even weakly zero. This problem could  be circumvented
by considering\footnote{Note the change of sign in the prefactor.}
\be
H' = \left(1+ \bar\xi\xi\right) \left| \tilde p +i\kappa \bar z +i\xi
\left(\pi -\kappa\bar\zeta\right)\right|^2 ,
\ee
which has weakly vanishing Poisson brackets with the constraints
and is weakly equal to $H$.
However, this has the disadvantage that $H'$ depends on the fermionic momenta.
We prefer to proceed differently. We define the new anticommuting variables
\be
\xi^1 = \zeta + z\xi\, ,\qquad \xi^2 = \xi\, ,\label{Second}
\ee
and let $(\chi_1,\chi_2)$ be their canonically conjugate momenta. Defining
\be
p= \tilde p + i \xi\pi\, ,
\ee
we find that the Lagrangian in the new variables is
\be
L=  \left[\dot z p  -i \dot \xi^i \chi_i   + \lambda^i\varphi_i \right] + c.c. - H\,,
\ee
where $\lambda^i$ are Lagrange multipliers for the constraints 
$\varphi_i \approx 0$ $(i=1,2)$.
The constraint functions are
\bea
\varphi_1 &=& \chi_1 -\kappa\bar\xi_1\left(1-\bar\xi_2\xi^2\right) 
+ i \bar\xi_2 p\, , \nonumber\\
\varphi_2 &=& \chi_2 + \kappa z\bar\xi_1 \left(1-\bar\xi_2\xi^2\right) 
-i\bar\xi_2 zp - M \bar\xi_2\, ,
\eea
and the Hamiltonian  is now
\be
H = \left(1+ \bar\xi^2\xi_2\right)\left| p +i\kappa \bar z
-i\kappa\xi^2\left(\bar\xi_1 -\bar z\bar\xi_2\right) \right|^2.
\ee
This Hamiltonian has (strongly) vanishing Poisson brackets
with the constraints.  As before, all these constraints are second class
except on the surface $H=4\kappa M$.

\subsection{Quantization}
\label{sec:quant}

We will quantize the planar superflag model of the previous section using the Gupta-Bleuler
method;  details and references can be found  in our previous
papers \cite{Ivanov:2003ax,Ivanov:2003qq,Ivanov:2004yw}. This is a method
of quantization in the presence of  analytic  constraints that are second class only
when considered in conjunction with their complex conjugates, exactly as we found
for the constraints of the planar
superflag model. We also found that there is a surface on which these constraints are not second
class, but we will deal with this problem when and where it presents a difficulty.
We also work with the variables $(z,\xi^1,\xi^2)$ in this section.

The method instructs us to quantize initially as there were no constraint, so we make the usual
replacements
\be
p \to \hat p = -i\partial_z \,, \qquad \bar p \to -i\partial_{\bar z}\, , \qquad
\chi_i \to \hat\chi_i = \partial_{\xi^i}\, , \qquad \bar\chi^i = \partial_{\bar\xi_i}\, .
\ee
The Hamiltonian can be written in terms of the operators
\be
\nabla_z = \partial_z - \kappa \bar z + \kappa\xi^2
\left(\bar\xi_1 - \bar z \bar\xi_2\right) \, ,\qquad
\nabla_{\bar z} = \partial_{\bar z} + \kappa z - \kappa\bar\xi_2\left(\xi^1-z\xi^2\right),
\label{defNabla}
\ee
which satisfy
\be
[\nabla_z,\nabla_{\bar z}] = 2\kappa\left(1-\bar\xi_2\xi^2\right).\label{identity}
\ee
There is an operator ordering ambiguity in the quantum Hamiltonian,
but this affects only the choice of ground state energy. If we resolve
this ambiguity in the usual way we arrive at the Hamiltonian operator
\be
\hat H = -\frac12 \left(1+\bar\xi_2\xi^2\right) \{\nabla_z, \nabla_{\bar z}\} =
-\left(1+\bar\xi_2\xi^2\right) \nabla_z \nabla_{\bar z} + \kappa\,. \label{hatH}
\ee
This operator $\hat H$ is positive definite. As we shall shortly see, 
the lowest eigenvalue of
$\hat H$ is $\kappa$, so the cancellation of vacuum energies that we noted 
for the superplane
model no longer occurs. This is because the Hamiltonian no longer depends on $\zeta$. 
This raises
a puzzle because the vacuum energy of the $M=0$ planar superflag 
model is also equal to $\kappa$,
but this model is classically equivalent to the superplane model. 
There is thus an apparent
quantum inequivalence of the $M=0$ planar superflag model with 
the superplane Landau model, but
this is a trivial difference that could be removed by 
a different operator ordering prescription.
As we shall see, the equivalence holds quantum mechanically 
in all other respects.

The constraints are now taken into account by the physical state conditions
\be
\hat{\bar\varphi}^i \Psi =0\qquad (i=1,2)\,, \label{Statecond}
\ee
where
\bea
\hat{\bar\varphi}^1 &=& \partial_{\bar\xi_1} 
- \kappa\xi^1 \left(1-\bar\xi_2\xi^2\right) -
\xi^2\partial_{\bar z}\, , \nonumber\\
\hat{\bar\varphi}^2 &=& \partial_{\bar\xi_2} 
+ \kappa \bar z \xi^1 \left(1-\bar\xi_2\xi^2\right) +
\xi^2\bar z\partial_{\bar z} -M \xi^2\,.
\eea
Solving these constraints one finds that physical wavefunctions have the form
\be
\Psi =  K \,  \Phi\left( z, \bar z_{sh}, \xi^1,\xi^2\right)\, , \qquad \bar z_{sh}
= \bar z - \xi^2\left(\bar\xi_1 - \bar z\bar\xi_2\right), \label{Phys1}
\ee
where $K$  is a real prefactor which we write as
\be
K= K_1^M\, e^{-\kappa K_2}
\ee
with
\be
K_1= \left(1+\bar\xi_2\xi^2\right)\, ,\qquad
K_2 = \left[ |z|^2 + \left(\xi^1-z\xi^2\right)
\left(\bar\xi_1 -\bar z\bar\xi_2\right)\right].
\ee
Thus, physical states are described by ``chiral '' wavefunctions
$\Phi\left( z, \bar z_{sh}, \xi^1,\xi^2\right)$
(we use this term because of the close analogy
to chiral superfields in supersymmetric field theories). Observe that
\be
\nabla_{\bar z} \Psi = K \partial_{\bar z}\Phi\, , \qquad
\nabla_z \Phi = K \, \tilde \nabla_z \Phi \, ,
\ee
where
\be
\tilde\nabla_z = \partial_z - 2\kappa \bar z_{sh}\, .
\ee
This derivative has the property that it  preserves chirality by taking
a chiral wavefunction to another chiral wavefunction. It follows
that the differential operators $(\nabla_z,\nabla_{\bar z})$ become
the differential operators $(\tilde\nabla_z, \partial_{\bar z})$ 
in the chiral basis,
i.e., when acting on reduced wavefunctions.
In particular the hamiltonian operator $\hat H$ is replaced by
\be
\hat H_{red} = -  K_1  \tilde\nabla_z \partial_{\bar z} + \kappa
\ee
in the chiral basis.

Reduced ground state wavefunctions, of energy $kappa$, are {\it analytic}, so
ground state wavefunctions have the form
\be
\Psi^{(0)} = K\, \Phi^{(0)}_0(z,\xi^1,\xi^2)\, .
\ee
One can now generate an infinite set of eigenvectors of
$\hat H$ by considering:
\bea\label{state}\Psi^{(N)} =
\nabla_z^N \left[K
\Phi_0^{(N)}\left(z, \xi^i\right)\right]
                   =  K
                   \tilde\nabla_z^N \Phi_0^{(N)}\left(z,
\xi^i\right). \lb{neigen} \eea
Indeed, using the commutation relation
\be
\left[\partial_{\bar z}, \tilde\nabla_z ^N\right] =
-2\kappa N K_1^{-1} \tilde\nabla_z ^{N-1}\, ,
\ee
it can be seen that
\be
\hat H_{red}\left(\tilde\nabla_z^N \Phi^{(N)}_0\right)
= 2\kappa 
\left(N + \frac12\right)\left(\tilde\nabla_z^N \Phi^{(N)}_0\right), \label{Levels}
\ee
and hence that the wavefunctions \p{neigen}
are eigenfunctions of $\hat{H}$ with energy $2\kappa \left(N +\frac12\right)$.
Note that $\tilde\nabla_z$ preserves chirality, but not the analyticity,
so the reduced function $\Phi^{(N)}
= \tilde\nabla_z^N \Phi^{(N)}_0(z, \xi^1, \xi^2)$ is a particular case of $\Phi$ defined
in \p{Phys1}, with a special dependence on $\bar z_{sh}\,$. Note also that
the analytic ``ground state'' functions $\Phi^{(N)}_0$ for different $N$ differ in their
``external'' $C$ charge $\tilde{C} = 2M -N\,$. The wavefunctions $\Psi^{(N)}$ and $\Phi^{(N)}$
have the fixed charge $\tilde{C} = 2M$ for any $N$, since $\nabla_z$ and $\tilde\nabla_z$
carry $\tilde{C} = 1\,$ (see subsection 3.3).

We have now found the energy eigenstates so we turn to the question of their norm.
The integration measure
\be
d\mu = dz d{\bar z} \partial_{\bar\xi_1}\partial_{\xi^1} \partial_{\bar\xi_2}\partial_{\xi^2}
\ee
is invariant under the symmetries of the model established previously, so we define
the norm of $\Psi$ by
\be
|||\Psi||| ^2 = \int d\mu \, |\Psi|^2 = \int d\mu \, K_1^{2M}
 e^{-2\kappa K_2}\,  |\Phi|^2\, .
\ee
For a ground state, the reduced wavefunction is analytic and can be expanded as
\be
\Phi^{(0)}_0  = A^{(0)} + \xi^i\psi^{(0)}_i+ F^{(0)} \xi^1\xi^2\,,
\ee
where all the coefficients are functions of $z\,$. A calculation shows that its norm is
\be\label{normzero}
|||\Phi^{(0)}_0|||^2 = 4\kappa M||A^{(0)}||^2 + 2M || \psi_1^{(0)}||^2 +
2\kappa || \psi_2^{(0)} + z\psi_1^{(0)}||^2 + ||F^{(0)}||^2\, ,
\ee
where
\be
|| f ||^2 = \int dz d\bar z\, e^{-\kappa |z|^2}\, |f(z,\bar z)|^2
\ee
for any function $f$ on the complex plane. Note that we have a shortened multiplet
when $M=0$ because there are then states with zero norm. This is
the quantum manifestation of the classical observation that for
$M=0$ the constraints are not all second class when $H=0\,$.

Consider now the first excited states, at $N=1\,$. Integrating by parts with respect
to $\partial_z, \partial_{\bar z}$, one sees that
\be
|||\Psi^{(1)}|||^2 = 2\kappa \int d\mu\, K_1^{2M-1} e^{-2\kappa K_2} |\Phi^{(1)}_0|^2\, .
\ee
In other words, the coefficient $M$ is shifted downwards by $1/2$. Similarly,
\be
|||\Psi^{(N)}|||^2 = (2\kappa)^N N! \int d\mu\, K_1^{2M-N} e^{-2\kappa K_2} |\Phi^{(N)}_0|^2\, ,
\ee
so the coefficient $M$ is shifted downwards by $N/2$ at level $N$. It follows that
$|||\Psi^{(N)}|||$ is also given by the formula \p{normzero}, apart from the numerical 
factor $(2\kappa)^N N!$, but with $2M \rightarrow 2M -N$. Thus, negative
contributions to the norm must appear for $N>2M$. If $2M$ is a positive
integer then the highest level without negative norm states is the $(2M+1)$th
level with $N=2M$, but this level has zero norm states, as for $M=0$. The states
at this level will therefore form short supermultiplets as only the components
$\psi^{(2M)}_2 + z\psi^{(2M)}_1\,, \; F^{(2M)}$ contribute to $|||\Psi^{(N=2M)}|||\,$. 
The energy of the  $N=2M$ level for integer $2M$ is $4\kappa M + \kappa$. Apart 
from the quantum shift by $\kappa$ noted earlier, this is just the energy of the
exceptional energy surface $H = 4\kappa M$ of the classical theory. Zero norm 
states in the quantum theory at this level are what one expects from the fermionic gauge 
invariance at this level. 

Just as one can discard all excited states of the supersphere, or superplane, Landau model
to arrive at a perfectly physical LLL model, so we can discard 
all states in the $N>2M$ Landau levels 
of the superflag, or planar superflag, models  to arrive at a physical model described 
by the LLL together with the first $N$ excited levels. 
This remains true when $2M$ is 
not an integer (provided it is positive), the only difference being that the top level, with 
$N=[2M]\,$, has no zero norm states. 

\subsection{Geometrical interpretation}
\label{sect:GI}

So far we have  used a direct algebraic analysis because our aim has been
to show how the results of our previous paper on the superflag Landau model
can be understood very explicitly in the planar limit, without any elaborate formalism.
However, we now develop a geometrical interpretation in terms of superfields 
on the coset superspace
\be
{\cal K} = IU(1|1)/[U(1)\times U(1)\times {\cal Z}]\,. \label{Coset}
\ee
Recall that ${\cal Z}$ is the group generated by the ``magnetic'' central
charge $Z$, which we identify with the constant $\kappa\,$.

The coset representative in the appropriate exponential parametrization can be
written in terms of coordinates $(u,\eta^1,\eta^2)$ as
\be\label{coset}
g = e ^{{\cal A}_1} e^{{\cal A}_2}\,,
\ee
where\footnote{We take the Grassmann-odd coordinates $\eta^i$ to anticommute 
with the odd charges.
One can equally well take them to commute with the odd charges because 
with an appropriate
change in the definition (\ref{coset}) one obtains identical results.}
\be
{\cal A}_1 = \eta^1\Pi -\eta^2Q + \bar\eta_1\Pi^\dagger -\bar\eta_2 Q^\dagger\,, \quad
{\cal A}_2  = -i u P -i \bar u P^\dagger \, ,
\ee
where the signs are chosen for later convenience. The coordinates appearing in the above
parametrization of the coset superspace are related to the coordinates 
$(z,\zeta,\xi)$ used
previously by
\be\label{first}
u = z - {1\over 2}\zeta\bar\xi \,, \qquad
\eta^1= \zeta + z\xi -{1\over3}\bar\xi\xi\,, \qquad
\eta^2 = \xi\,.
\ee

The left-covariant Cartan forms and the superconnections on the stability  subgroup
generated by $C$ and the central charge $\kappa$ are defined
by\footnote{As the second $U(1)$ in the denominator of \p{Coset} corresponds to
an outer automorphism of  $ISU(1|1)$ (see \p{Jdef}, \p{Jcomm}), there appears
no connection associated with its generator $J$.}
\be
g^{-1} dg = i\omega_P P + i\bar\omega_P P^\dagger + \omega^1 \Pi 
+ \bar\omega_1 \Pi^\dagger -
\omega^2 Q - \bar\omega_2 Q^\dagger + A_C C + A_{2\kappa} \kappa\,. \label{CaF}
\ee
A calculation yields\footnote{The
$A_{2\kappa}$ connection given here is equivalent to the connection defined by \p{CaF}
but differs from it by a field-dependent gauge transformation.}
\bea
&& \omega_P = - \left(1+ {1\over2}\bar\xi\xi\right)dz - \bar\xi d\zeta\, , \qquad
\omega^1 = \xi dz + \left(1- {1\over2}\bar\xi\xi\right) d\zeta\,,
\qquad \omega^2 = d\xi\,, \nonumber\\
&& A_{2\kappa} = -\left( \bar z\,dz -  z\,d {\bar z}  - \bar\zeta\, d \zeta 
- \zeta\, d {\bar\zeta}\right), \quad
A_C = \frac12\left(\xi d\bar\xi + \bar\xi d \xi\right). \label{Conn}
\eea

It is now easy to rewrite the invariant Lagrangians \p{L0}, \p{altlag} and \p{planarSF}
of the previous sections in a manifestly invariant  form in terms 
of pullbacks of the above Cartan forms:
\be\label{LagR}
L_0 = |\hat\omega_P|^2 + \hat\omega^1\hat{\bar\omega}_1 +i\kappa\hat{A}_{2\kappa}\,, \quad
\tilde{L} = |\hat\omega_P |^2 + i\kappa\hat{A}_{2\kappa}\,, \quad
L = |\hat\omega_P |^2 + i\kappa\hat{A}_{2\kappa} + 2iM \hat{A}_C\, .
\ee
Here the ``hat''  denotes a pullback. Note that the passage from the superplane Landau model,
with Lagrangian $L_0$, to the $M=0$ planar superflag model, with Lagrangian $\tilde L$,
involves the subtraction  of the term $\hat\omega^1\hat{\bar\omega}_1$. The Lagrangian $L_0$ is
necessarily independent of the $\xi, \bar\xi$ variables because it is invariant under
local $SU(1|1)$ transformations that rotate the forms $\omega_P$ and $\omega^1$ 
(and their conjugates) into each other.

Note also that the equation of motion \p{EqKsi} derived from $\tilde{L}$ has the
following nice representation in terms of the Cartan forms:
\be
\hat\omega^1 \hat{\bar\omega}_P = 0\,. \label{EqCF}
\ee
This equation has two solutions. One is
\be
\hat\omega^1 = 0\,,
\ee
which a covariant inverse Higgs-type constraint \cite{Ivanov:1975zq} that is
equivalent to  \p{Ksi}. The other is
\be
\hat\omega_P = 0\quad \Rightarrow \quad \dot z = -\bar\xi \dot\zeta\,,   \label{Altern}
\ee
in which case all other equations of motion are identically satisfied. As we have seen,
this second solution reduces the model to its LLL sector.

Finally, we explain the geometric meaning of the wavefunctions $\Psi^{(N)}$ which are
eigenvectors of the Hamiltonian $\hat{H}$ defined in \p{hatH}.
As a first step, we note that the full generators $\hat Q$, $\hat{Q}^\dagger$
calculated by the Noether procedure from the Lagrangian $L$ defined in \p{planarSF} are
given by
\be
\hat Q = Q - \frac{\partial}{\partial \xi} - M \bar\xi\,,
\quad \hat{Q}^\dagger  = Q^\dagger- \frac{\partial}{\partial \bar\xi} - M \xi\,,
\ee
where $Q, Q^\dagger$ were defined in \p{QQbar}. Correspondingly, the full $C$ charge
appearing in $\{\hat Q, \hat Q^\dagger\} = \hat C$ is given by
\be
\hat C = C + 2M \equiv C + \tilde{C}\,,
\ee
where $C$, the purely differential part of $\hat C$, was defined in \p{Ccharge}.
The additional term $\tilde{C} = 2M$ can be interpreted as the 
``external'' $C$ charge of the
general wavefunction $\Psi(z,\bar z, \xi, \bar\xi, \zeta, \bar\zeta)$, 
in accordance with the
fact that this function is given on the coset manifold 
$IU(1|1)/[U(1)\times U(1)\times {\cal Z}]$
and can possess non-zero
quantum numbers of the stability subgroup. The generator $Z$ acts on $\Psi$ just
as the multiplication of the latter by the central 
charge $\kappa$.\footnote{One can assign to
$\Psi$ also a non-zero external charge associated with the outer 
automorphisms $U(1)$ generator
$J$ the differential part of which is given in \p{Jdef}. However, 
this $U(1)$ has no
actual implications in the considered model.} Thus 
the wavefunction $\Psi$ carries the
``magnetic'' central charge $\kappa$ and the 
external $C$ charge $\tilde{C} = 2M$.

For the next step we find it convenient to use the parametrization 
$(z, \bar z, \xi^i, \bar\xi_i)$
of subsection \ref{sec:quant}.  In accord with the standard 
rules of the nonlinear realizations theory,
the covariant differential ${\cal D}\Psi$ of $\Psi$, 
as well as covariant derivatives of $\Psi$ are defined by
the relation
\be
{\cal D}\Psi = \left(d + A_{2\kappa}\,\kappa  +  A_C\,\tilde{C}\right)\Psi
\equiv -\omega_P {\cal D}_z\Psi - \bar\omega_P{\cal D}_{\bar z}\Psi
+ \omega^i{\cal D}_i\Psi + \bar\omega_i \bar{\cal D}^i\Psi\,,
\ee
where the signs were again chosen for further convenience. It is easy to
find the explicit form of these covariant derivatives. In
particular,
\be
{\cal D}_z = K_1^{\frac12} \nabla_z\,,
\quad {\cal D}_{\bar z} =
K_1^{\frac12} \nabla_{\bar z}\,,
\quad \{{\cal D}_z, {\cal D}_{\bar z} \}
= 2\kappa\,, \label{ident2} \ee
where $\nabla_z, \nabla_{\bar z}$
were defined in \p{defNabla}.
The covariant spinor derivatives $\bar{\cal D}^i$ are:
\bea \bar{\cal D}^1 = K_1^{\frac12}\left({
\partial \over {\partial{\bar \xi}_{1}}} -\xi^2\partial_{\bar z} -
\kappa \xi^1\,K_1^{-1}\right)\,, \quad \bar{\cal D}^2 = {\partial
\over {\partial{\bar \xi}_{2}}} + \bar z {
\partial \over {\partial{\bar \xi}_{1}}} - \frac12\xi^2\,\tilde{C}. \label{CovD}
\eea
They satisfy the following non-zero covariant (anti)commutation relations
\bea
&&[\bar{\cal D}^1, {\cal D}_z] =[\bar{\cal D}^1, {\cal D}_{\bar z}] = 0\,, \quad
[\bar{\cal D}^2, {\cal D}_z] =0\,, \quad [\bar{\cal D}^2, {\cal D}_{\bar z}]
= - \bar{\cal D}^1\,, \label{DD1} \\
&&\{\bar{\cal D}^1, \bar{\cal D}^2\} = 0\,. \label{DD2}
\eea
One should take into account that all coset coordinates
and their covariant
derivatives are inert under the action of the ``magnetic''
central charge $Z$ which has
the non-zero eigenvalue $\kappa$ only on the wave function
$\Psi$; at the same time, the $U(1)$ charge
$C$ has a non-trivial left action on the coset coordinates
$z, \bar z, \xi^1, \bar\xi_1$
as follows from the commutation relations \p{102}. Under the
above normalization, such that $\Psi$ has the
external $\tilde C$ charge equal $2M$, the covariant derivatives
$\bar{\cal D}^1\,$,
${\cal D}_z\,$, ${\cal D}_{\bar z}$ have, respectively, the
$\tilde{C}$ charges $+1, +1$ and
$-1$, while ${\cal D}^2, \bar{\cal D}_2$ are $\tilde C$-neutral.
This $\tilde C$ assignment should be kept in mind while checking
the relations \p{DD1}, \p{DD2}. The
standard (non-covariant) commutation relations (without taking
account of the non-trivial
$\tilde C$ connection terms in ${\cal D}_2, \bar{\cal D}^2$)
can be easily derived from the above
covariant ones.

Representing the covariant derivatives $\bar{\cal D}^i$ on $\Psi$ 
(i.e. with $\tilde{C} = 2M$) by
\be
\bar{\cal D}^1 = K^{\frac12}_1\bar\varphi^1\,, \quad \bar{\cal D}^2
= \bar\varphi^2 + \bar z \bar\varphi^1\,,
\ee
it is easy to see that the physical state conditions \p{Statecond}
are equivalent to
\be \bar{\cal D}^i \Psi = 0\,, \lb{Chir}
\ee
which is the standard covariant form of the chirality conditions. The prefactors in
the solution \p{Phys1} serve to eliminate the connection terms in $\bar{\cal D}^i$
when the latter
act on the reduced wave function $\Phi\,$. After that, the conditions \p{Chir}
are solved by passing
to the chiral basis $(z, \bar z_{sh})$. The derivative ${\cal D}_{\bar z}$
also becomes short on
$\Phi(z, \bar z_{sh}, \xi^i)$: ${\cal D}_{\bar z}\; 
\rightarrow \;\tilde{\cal D}_{\bar z}
= K^{\frac12}_1\partial_{\bar z_{sh}}$. Thanks to the commutation relations \p{DD1},
it is
then consistent to impose the additional analyticity constraint on the ground state
$\Phi(z, \bar z_{sh}, \xi^i)$, viz.
$\tilde{\cal D}_{\bar z}\Phi = 0 \; \rightarrow \;\Phi = \Phi_0 (z, \xi^i)\,$.

When dealing with the eigenvalue problem of the Hamiltonian in the previous subsection,
we worked with the operators $\nabla_z\,, \nabla_{\bar z}$, which
can be treated as a type of creation and annihilation operator (see \p{identity}).
Using the covariant derivatives ${\cal D}_z, {\cal D}_{\bar z}$, eq. \p{ident2},
the analogy with
the quantum oscillator becomes literal, because their commutator equals
a constant and the Hamiltonian can be rewritten in the standard oscillator form:
\bea\label{hamiltonian3}
\hat H =
-{\cal D}_z {\cal D}_{\bar z} + \kappa\,.
\eea
The eigenvector for the Landau level $N$ can be rewritten
as
\bea\label{state1}\Psi^{(N)} =
\left({\cal D}_z\right)^N K_1^{M -\frac{N}{2}} e^{-\kappa K_2}
\Phi^{(N)}_0\left(z, \xi^i\right).
\eea
The corresponding ground state reduced wave function $\Phi^{(N)}_0$ 
has $\tilde{C} = 2M - N\,$,
while the whole $\Psi^{(N)}$ has $\tilde{C} = 2M$, since each ${\cal D}_z$
adds $\tilde{C} = 1$. The formula \p{Levels} for the energy levels can be
equivalently derived using the commutation relations \p{ident2}. Note that
the Hamiltonian commutes with the chirality constraints \p{Chir} in a weak sense,
$[\hat H, \bar{\cal D}^2] \sim \bar\varphi^1\,$.

\section{Summary}
\setcounter{equation}{0}

In previous papers we solved the Landau problem for a particle on the
supersphere $SU(2|1/U(1|1)$ and the superflag  $SU(2|1)/[U(1)\times U(1)]$. The latter
coset superspace allows two WZ terms, and hence a family of Landau models, for fixed
magnetic field, parametrized by the coefficient $M$ of a ``fermionic Wess-Zumino''  term.
The equivalence of the $M=0$ model with the supersphere Landau model was implicit
in these results, but not explained by them. In this paper we have reconsidered these
models in the planar limit.

The supersphere model becomes the ``superplane'' Landau model
for a particle on $\bC^{(1|1)}$; this is a  model with a quadratic Lagrangian
that is the sum of the standard Landau model with a four-state ``fermionic Landau model''.
The latter has just two Landau levels, each spanned by two states, 
with the excited states
having negative norm.
This provides a simple explanation for the negative norm states, or ``ghosts'',
in all but the lowest
Landau level of the supersphere model, and it shows clearly that ghosts arise
as a result of second-order fermion kinetic terms.

The planar limit of the superflag model yields a model that we have called
the ``planar superflag'' Landau model. It is an extension of the superplane
to include interactions with
an additional Goldstino variable.  For $M=0$ this variable is auxiliary and
the superplane model is recovered on eliminating it; this explains the equivalence
between the superplane and
$M=0$ superflag models. The motivation for considering the $M>0$ superflag model
(planar or spherical) is that the second-order fermion kinetic terms responsible
for ghosts are ``suppressed'' in the sense that the coefficient becomes nilpotent.
As a result, the ghosts are not eliminated entirely but just banished
to the higher Landau levels. Specifically, the $N$th level is ghost-free 
if and only if $N\le2M$.

Another curious, and related, feature of the $M>0$ planar superflag models
is that  the second class  fermionic  constraints (which are standard
in models with anticommuting variables) cease to be entirely second-class
on a fixed-energy subspace of the phase space, thus implying the presence
of a gauge-invariance on this energy surface. In fact, when restricted to this
exceptional energy the planar superflag Landau model becomes a type of
time-reparametrization invariant superparticle model with a ``hidden'' fermionic
gauge invariance. However, this gauge invariance has an effect on the quantum
theory only when the exceptional energy surface is one of the Landau levels, and this
happens only when $2M$ is an integer. In this case, the fermionic gauge invariance
leads to short supermultiplets for the states at the $(2M+1)$th Landau level, 
this being the lowest Landau level for $M=0$. The short supermultiplets are exactly
as expected from our previous results for the supersphere and superflag
Landau models.

Although the super-Landau models analysed  here have ghosts, it is possible to
consistently truncate to a ghost free theory. One could throw out just the ghosts, but this
would break the $SU(2|1)$ symmetry that was the rationale for the construction of 
these models. Instead, one can throw out all Landau levels that contain ghosts. 
For $M=0$ this is
equivalent to keeping only the lowest Landau level,  which defines 
the non-(anti)commutative
complex superplane that results from taking the planar limit of the fuzzy  supersphere. 
Our $M>0$
planar superflag models, truncated to the first $2[M]+1$ levels,  can be considered as
generalizations of this construction to allow for a finite set of higher Landau levels. As
the Hilbert space still has finite dimension, the quantum theory  defines a fuzzy version of
the supermanifold obtained from the planar limit of the superflag.

\subsection*{Note Added}

After submission to the archives, we learnt of a paper of Hasebe 
\cite{hasebe} in which a planar super-Landau model is obtained as
the planar limit of a Landau model for a particle on the coset
superspace $OSp(1|2)/U(1)$. This ``supersphere'' has real dimension 
$(2|4)$, and is
therefore ``non-minimal'' in comparison to the supersphere defined here 
as $CP^{(1|1)}$, but it can be viewed as a superspace of real dimension
$(2|2)$ with the help of a ``pseudoconjugation'' operation that
squares to $-1$ when acting on spinors. This leads to a planar
super-Landau model that is superficially equivalent to the superplane
Landau model discussed here, but which has a different symmetry group.
The absence of negative norm states in the model of \cite{hasebe}
is presumably a consequence of this difference. We believe that
the consistency of the Hilbert space norm of \cite{hasebe} 
requires an interpretation as a ``bi-orthogonal'' norm \cite{bender}
(see also \cite{Curtright:2005zk}), and we plan to return to this 
point in a future work with T. Curtright.

\setcounter{section}{0}

\subsection*{Acknowledgments}
E.I. acknowledges a partial support from RFBR
grants, projects  No 03-02-17440 and No 04-02-04002, NATO grant
PST.GLG.980302, the grant INTAS-00-00254, the DFG grant No.436 RUS
113/669-02, and a grant of the Heisenberg-Landau program. 
Part of this paper was presented at the Miami 
2004 topical conference on particle physics and cosmology. 
L.M. and P.K.T. thank the organisers for the invitation to participate in
the 2005 Strings workshop at Benasque, where some of this work was
done. In addition, we thank C. Bender, T. Curtright, M. Henneaux 
and A. Smilga for helpful
discussions, and K. Hasebe for bringing his work to our attention.

\end{document}